\documentclass[twocolumn,amsmath,amssymb,pre]{revtex4}

\usepackage{graphicx}
\usepackage{bm}
\usepackage{amssymb}
\usepackage{amsmath}
\usepackage{amsfonts}

\usepackage[utf8]{inputenc}

\begin{document}

\title{Self-consistent equilibrium of a force-free magnetic flux rope}

\author{Oleg K. Cheremnykh}
\affiliation{Space Research Institute, Pr. Glushkova 40 k.4/1,
Kyiv 03187, Ukraine}

\author{Volodymyr M. Lashkin}
\email{vlashkin62@gmail.com} \affiliation{$^1$Institute for
Nuclear Research, Pr. Nauki 47, Kyiv 03028, Ukraine}
\affiliation{$^2$Space Research Institute, Pr. Glushkova 40 k.4/1,
Kyiv 03187,  Ukraine}

\begin{abstract} We present an exact solution
to the problem of a self-consistent equilibrium force-free
magnetic flux rope. Unlike other approaches, we use magnetostatic
equations and assume only a relatively rapid decrease in the axial
magnetic field at infinity. For the first time we obtain a new
nonlinear equation for the axial current density, the derivation
of which does not require any phenomenological assumptions. From
the resulting nonlinear equation, we analytically find the radial
profiles of the components of the magnetic field strength and
current density.
\end{abstract}

\maketitle

The concept of a magnetic flux rope is widely used in solar
physics and has been an area of active research over the years. In
particular, it is assumed that such structures play a dominant
role in the release of flare energy and in the formation of
coronal mass ejections
\cite{Ryutova2015,Priest2000,Daughton2006,Chen2017,Aschwanden2019}.
Magnetic clouds, which are large-scale magnetic ropes, are also
often observed in interplanetary space
\cite{Krimigis1976,Klein1982}. As a theoretical model, the
magnetic ropes are usually considered as plasma tubes with a
helical magnetic field. Among these models, much attention has
been paid to the so-called force-free plasma configurations, in
which the current is parallel to the magnetic field. It is
believed \cite{Woltjer1958} that in closed plasma systems a
force-free magnetic field is a field with minimal energy and
therefore the force-free magnetic ropes are stable entities
widespread in space plasma.

The simplest force-free configurations usually correspond to
cylindrical symmetry, and then the force-free plasma tube is
considered in a cylindrical coordinate system $(r,\varphi,z)$ as a
plasma cylinder with two components of magnetic field strength
$B_{\varphi}$ and $B_{z}$. Since within the framework of such a
model there is only one equation for the magnetic field components
(see, e.g., \cite{Priest2000}), there is an arbitrariness in the
choice of configurations. The most well-known and frequently used
force-free cylindrical magnetic configurations are those obtained
by Lundquist \cite{Lundquist1951,Bothmer1999,Miyamoto2005} and
Gold-Hoyle \cite{Gold1960,Farrugia1999,Allanson2016}. The
force-free magnetic fields of Lundquist and Gold-Hoyle are exact
solutions to the equations of plasma hydrostatics, however, they
are not the only solutions within the framework of the force-free
plasma cylinder model and the problem of finding other solutions
is still ongoing
\cite{Marubashi2007,Vandas2017,Wang2018,Vandas2019}. It should be
noted that the results obtained by Lundquist \cite{Lundquist1951}
and Gold-Hoyle \cite{Gold1960} were obtained under quite
significant additional restrictions. In \cite{Lundquist1951} it
was assumed that the magnetic field strength and current density
are related by a constant value, and in \cite{Gold1960} a constant
angle of twist of the magnetic field lines was assumed. That is,
the forceless equilibrium in those works was "set by hand".

In this paper, we do not make any additional assumptions about the
behaviour of equilibrium quantities in force-free configurations
and use the approach of \cite{Cheremnykh2023}, which allows us to
analytically find a self-consistent equilibrium. Following this
paper, we find and add an integral equilibrium equation to the
three commonly used local equilibrium equations. We obtain a new
nonlinear equation for the axial current density and this makes it
possible to determine uniquely the self-consistent equilibrium
state of a force-free magnetic rope.

We consider an inhomogeneous equilibrium state of a force-free
plasma configuration, which is established self-consistently when
a helical current flows in a low-pressure plasma. Since in the
configuration under consideration the Lorentz force significantly
exceeds the pressure gradient, the standard equations of magnetic
hydrostatics, in which we neglect the gas-kinetic pressure,
\begin{equation}
\mathbf{j}\times\mathbf{B}=0, \label{max1}
\end{equation}
\begin{equation}
\nabla\times\mathbf{B}=\frac{4\pi}{c}\mathbf{j}, \label{max2}
\end{equation}
\begin{equation}
\nabla\cdot\mathbf{B}=0, \label{max3}
\end{equation}
are a good approximation for describing the equilibrium of a
force-free magnetic rope \cite{Priest2000}. Throughout the paper,
cgs units  are used. Here $\mathbf{j}$ is the electric current
density, $\mathbf{B}$ is the magnetic field strength, and $c$ is
the speed of light. From equation (\ref{max1}) it follows that the
electric current flows along the magnetic field lines, and
therefore we can write
\begin{equation}
\label{mu} \mathbf{j}=\mu\mathbf{B},
\end{equation}
where unlike \cite{Lundquist1951} we do not assume the coefficient
$\mu$ to be a constant.

Let us assume that in a cylindrical coordinate system the
components of the magnetic field strength $\mathbf{B}$ and
electric current density $\mathbf{j}$ depend only on the radial
coordinate $r$, that is, we consider an axially symmetric
configuration with $\mathbf{B}=(0,B_{\varphi} (r),B_{z}(r))$ and
$\mathbf{j}=(0,j_{\varphi} (r),j_{z}(r))$. It can be seen that in
this case eq. (\ref{max3}) is satisfied automatically. Magnetic
field lines lie on cylindrical surfaces with radius $r$ and form
spirals with a twist of field lines per unit length of the plasma
cylinder $k$ equal to
\begin{equation}
\label{twist} k=\frac{d\varphi}{dz}=\frac{B_{\varphi}}{rB_{z}}.
\end{equation}
Equations (\ref{max1}) and (\ref{max2}) in a cylindrical
coordinate system have the form
\begin{equation}
\label{max11} j_{z}B_{\varphi}-j_{\varphi}B_{z}=0,
\end{equation}
and
\begin{equation}
\label{max22} j_{\varphi}=-\frac{c}{4\pi}\frac{dB_{z}}{dr}, \quad
j_{z}=\frac{c}{4\pi r}\frac{d}{dr}(rB_{\varphi}),
\end{equation}
respectively. Eliminating the current density from eqs.
(\ref{max11}) and (\ref{max22}), we have equation
\begin{equation}
\label{basic1}
\frac{1}{2}\frac{dB_{z}^{2}}{dr}+\frac{B_{\varphi}}{r}\frac{d}{dr}(rB_{\varphi})=0,
\end{equation}
containing two independent radial profiles $B_{z}(r)$ and
$B_{\varphi}(r)$. A unique solution to the equilibrium equations
of a magnetic flux rope can be obtained if one more equation for
equilibrium quantities is added to them. In a recent work
\cite{Cheremnykh2023} on the study of the equilibrium of a
magnetic flux rope with a finite plasma pressure with a constant
$B_{z}$, it was shown that such an equation can be found from the
integral equilibrium equation. For a force-free plasma
configuration, we similarly obtain an additional equilibrium local
equation using the integral equilibrium equation.

Rewriting eq. (\ref{basic1}) in the form
\begin{equation}
\label{basic2}
r^{2}\frac{dB_{z}^{2}}{dr}+\frac{d}{dr}(rB_{\varphi})^{2}=0,
\end{equation}
and integrating it over $r$ from $0$ to $\infty$, we have
\begin{equation}
\label{basic3}
\int_{0}^{\infty}\frac{dB_{z}^{2}}{dr}r^{2}dr=-\left.(rB_{\varphi})^{2}
\right|_{0}^{\infty}.
\end{equation}
Integrating eq. (\ref{max22}) one can obtain
\begin{equation}
\label{au1} \left.rB_{\varphi}
\right|_{r\rightarrow\infty}=\frac{2I_{z}}{c},
\end{equation}
where $I_{z}$ is the total current flowing through the plasma in
the $z$-direction:
\begin{equation}
\label{Iz} I_{z}=2\pi\int_{0}^{\infty}j_{z}\,r\,dr.
\end{equation}
Note that the regularity condition on $B_{\varphi}$ ensures that
$rB_{\varphi}\rightarrow 0$ at $r=0$. Equation (\ref{basic3}) then
becomes
\begin{equation}
\label{basic4}
\int_{0}^{\infty}\frac{dB_{z}^{2}}{dr}r^{2}dr=-\frac{4I_{z}^{2}}{c^{2}}.
\end{equation}
Integrating by parts the left side of this equation gives
\begin{equation}
\label{basic5}
\left.(rB_{z})^{2}\right|_{0}^{\infty}-2\int_{0}^{\infty}B_{z}^{2}\,r\,dr=-\frac{4I_{z}^{2}}{c^{2}}.
\end{equation}
Assuming that $B_{z}^{2}$ is bounded at $r=0$ and vanishes at
infinity faster than $\sim 1/r^{2}$, we obtain the following
integral equilibrium equation for a force-free magnetic flux rope:
\begin{equation}
\label{basic6}
\int_{0}^{\infty}B_{z}^{2}r\,dr=\frac{2I_{z}^{2}}{c^{2}}.
\end{equation}
Taking into account eq. (\ref{Iz}), one can rewrite eq.
(\ref{basic6}) in the form
\begin{equation}
\label{basic7} \int_{0}^{\infty}\left(B_{z}^{2}-\frac{4\pi
j_{z}I_{z}}{c^{2}}\right)r\,dr=0.
\end{equation}
It can be seen that equation (\ref{basic7}) is satisfied if the
$z$-components of the magnetic field strength $B_{z}$ and current
density $j_{z}$ satisfy the condition:
\begin{equation}
\label{alpha} B_{z}^{2}=\alpha j_{z},
\end{equation}
where $\alpha=4\pi I_{z}/c^{2}$ is a constant determined by the
total current. Setting $r=0$ in eq. (\ref{alpha}), the constant
$\alpha$ can also be written as $\alpha=B_{0}^{2}/j_{0}$, where we
have introduced notations for the values of the axial magnetic
field and current density on the axis:
\begin{equation}
\label{B0j0} B_{0}=\left.B_{z}\right|_{r=0}, \quad
j_{0}=\left.j_{z}\right|_{r=0}.
\end{equation}
It is assumed that the axial current density $j_{z}$ is non-zero.
This assumption is consistent with the subsequent explicit
expression for $j_{z}$. Then, the total current can be written as
\begin{equation}
\label{Iz-new}   I_{z}=\frac{c^{2}B_{0}^{2}}{4\pi j_{0}}.
\end{equation}
Equation (\ref{alpha}) is the desired additional equilibrium local
equation and, along with eqs. (\ref{max1})-(\ref{mu}), allows us
to self-consistently find all equilibrium quantities. From eqs.
(\ref{max22}), (\ref{basic1}) and (\ref{alpha}) we can obtain a
nonlinear equation for the axial current density $j_{z}$:
\begin{equation}
\label{nonlinear1}
\frac{1}{r}\frac{d}{dr}\left(\frac{r}{j_{z}}\frac{dj_{z}}{dr}\right)+\frac{8\pi}{I_{z}}j_{z}=0.
\end{equation}
The solution of this equation with boundary conditions
\begin{equation}
\label{boundary} j_{z}=j_{0} \quad \mathrm{at} \quad r=0, \quad
j_{z}\rightarrow 0 \quad \mathrm{at} \quad r\rightarrow\infty,
\end{equation}
along with eqs. (\ref{max22}), (\ref{basic1}) and (\ref{alpha})
allows us to self-consistently find all equilibrium quantities. As
will be shown below, this solution can be extended to the problem
where the current is mainly contained in a plasma cylinder with a
certain radius $a$. Such a cylinder can be considered as a
force-free magnetic flux rope. Let us represent the axial current
density in the form $j_{z}=j_{0}\exp \psi$, where the function
$\psi (r)$, according to (\ref{boundary}), satisfies the boundary
conditions $\psi=0$ at $r=0$ and $\psi\rightarrow -\infty$ at
$r\rightarrow\infty$. Then from eq. (\ref{nonlinear1}) we obtain a
nonlinear equation for the function $\psi$:
\begin{equation}
\label{nonlinear2}
\frac{1}{r}\frac{d}{dr}\left(r\frac{d\psi}{dr}\right)+\beta\exp\psi=0,
\end{equation}
where $\beta=32\pi^{2}j_{0}/(\alpha c^{2})$. The solution to
equation (\ref{nonlinear2}) was found in \cite{Cheremnykh2023} and
has the form:
\begin{equation}
\label{solution} \psi=\ln \frac{1}{(1+r^{2}/r_{\ast}^{2})^{2}},
\end{equation}
where
\begin{equation}
\label{r-star}
r_{\ast}^{2}=\frac{8}{\beta}=\frac{c^{2}B_{0}^{2}}{4\pi^{2}j_{0}^{2}}=\frac{I_{z}}{\pi
j_{0}}.
\end{equation}
Note that $r_{\ast}=2b$, where $b$ is the unit of length
introduced in the numerical simulation of force-free
configurations in \cite{Alfven-book}. The radial dependence of the
axial current density $j_{z}$ is readily found from
(\ref{solution}):
\begin{equation}
\label{jz} j_{z}=\frac{j_{0}}{(1+r^{2}/r_{\ast}^{2})^{2}}.
\end{equation}
The value $r_{\ast}$ can be treated as the radius of the cross
section of the cylinder, inside of which the total axial current
$I_{z}=\pi j_{0}r_{\ast}^{2}$ flows with a uniform current density
$j_{0}$. From eqs. (\ref{max22}), (\ref{basic1}) and (\ref{alpha})
we find for the azimuthal magnetic field:
\begin{equation}
\label{B-phi1} B_{\varphi}=-\frac{\alpha c}{8\pi
j_{z}}\frac{dj_{z}}{dr},
\end{equation}
and substituting (\ref{jz})  into (\ref{B-phi1}) we have:
\begin{equation}
\label{B-phi2}
B_{\varphi}=B_{0}\frac{(r/r_{\ast})}{(1+r^{2}/r_{\ast}^{2})},
\end{equation}
where
\begin{equation}
\label{B0} B_{0}=\frac{2\pi r_{\ast}j_{0}}{c}=\frac{2\sqrt{\pi
I_{z}j_{0}}}{c}.
\end{equation}
The axial magnetic field $B_{z}$ can be found from eq.
(\ref{alpha}) and (\ref{jz}):
\begin{equation}
\label{B-z} B_{z}=\frac{B_{0}}{(1+r^{2}/r_{\ast}^{2})}.
\end{equation}
Note that the $B_{0}$ defined by eq. (\ref{B0}) is consistent, as
can be seen from eq. (\ref{B-z}), with the boundary condition
(\ref{B0j0}). In turn, from eqs. (\ref{max22}), (\ref{B0}), and
(\ref{B-z}) we have for the azimuthal current density:
\begin{equation}
\label{j-phi}
j_{\varphi}=j_{0}\frac{(r/r_{\ast})}{(1+r^{2}/r_{\ast}^{2})^{2}}.
\end{equation}
Equations (\ref{jz}), (\ref{B-phi2}), (\ref{B-z}), and
(\ref{j-phi}) give expressions for the components of magnetic
field strength and current density for an equilibrium force-free
magnetic flux rope and present the main result of this Letter.
This force-free plasma configuration is analytically obtained from
the magnetostatic equations only under the assumption that the
axial magnetic field decrease at infinity faster than $\sim 1/r$.
It can be seen that all equilibrium quantities depend on two free
parameters $j_{0}$ and $B_{0}$ (or $j_{0}$ and $I_{z}$), which
determine the self-consistent equilibrium of the force-free
configuration of the magnetic flux rope. From the obtained
expressions it also follows that
$j_{\varphi}/j_{z}=B_{\varphi}/B_{z}=r/r_{\ast}$, so that the
magnetic field and current density at distances small compared to
$r_{\ast}$ have mainly $z$-components, and at large distances,
respectively, $\varphi$-components. Thus, the configurations of
magnetic field lines and  streamlines resemble the helical ropes.

From eqs. (\ref{twist}), (\ref{B-phi2}) and (\ref{B-z}) it follows
that the twisting of the magnetic field lines for the resulting
equilibrium magnetic field is equal to $k=1/r_{\ast}$. Up to the
change $1/r_{\ast}\rightarrow k$, the expressions for
$B_{\varphi}$ and $B_{z}$ coincide with the expressions for
components of the Gold-Hoyle magnetic field strength
\cite{Gold1960}. However, unlike the Gold-Hoyle solution, the
twist of the field lines for the resulting solution is not an
arbitrary constant value, but is expressed in terms of $j_{0}$ and
$I_{z}$, and  has the form:
\begin{equation}
\label{k} k=\sqrt{\frac{\pi j_{0}}{I_{z}}}=\frac{2\pi
j_{0}}{cB_{0}},
\end{equation}
where $k$ is a constant, but depends on the values of $j_{0}$ and
$I_{z}$. As follows from eqs. (\ref{jz}), (\ref{B-phi2}),
(\ref{B-z}), and (\ref{j-phi}), the coefficient $\mu$ in equation
(\ref{mu}) is not a constant but depends on $r$ and is equal to
\begin{equation}
\mu=\frac{j_{0}}{B_{0}}\frac{1}{(1+r^{2}/r_{\ast}^{2})}=\frac{c}{2}\sqrt{\frac{
j_{0}}{\pi I_{z}}}\frac{1}{(1+\pi j_{0}r^{2}/I_{z})}.
\end{equation}

\begin{figure}
\includegraphics[width=3.2in]{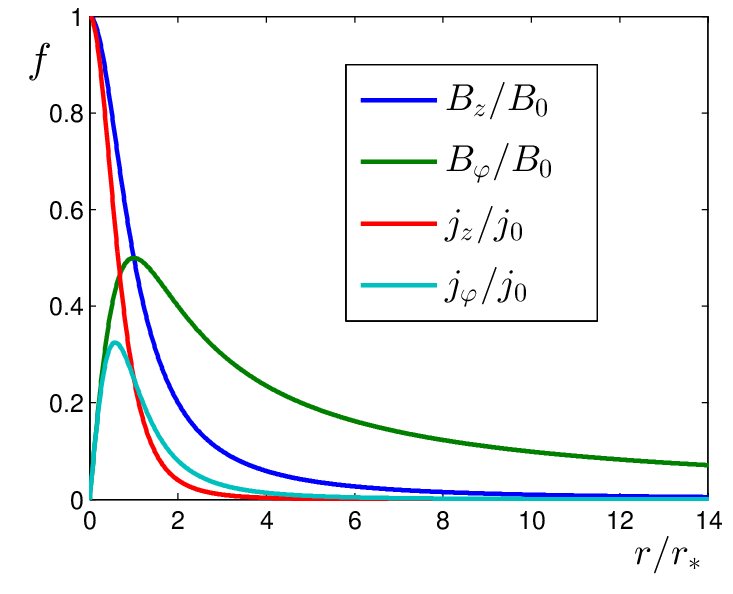}
\caption{\label{fig1} Radial dependences of the normalized
equilibrium quantities $f$: axial magnetic field $B_{z}$,
azimuthal magnetic field $B_{\varphi}$, axial current density
$j_{z}$, and azimuthal current density $j_{\varphi}$.}
\end{figure}

Radial profiles of equilibrium quantities $B_{\varphi}$, $B_{z}$,
$j_{\varphi}$, and $j_{z}$ are shown in Fig.~\ref{fig1}. Note that
they qualitatively coincide with the profiles obtained from the
numerical simulation in \cite{Alfven-book}. The axial magnetic
field  $B_{z}$ and current density $j_{z}$ decrease monotonically
with increasing $r$ and exceed the azimuthal magnetic field
$B_{\varphi}$ and current density $j_{\varphi}$ up to
$r=r_{\ast}$, and then become smaller compared to the azimuthal
magnetic field and current density. The azimuthal magnetic field
and current density increase from zero at $r=0$ to maximum values
at $r=r_{\ast}$ and $r\approx 0.577r$ respectively. Then at
$r=r_{\ast}$, they are equalized with the axial magnetic field and
current density and decrease monotonically.

Since the radial profiles of the found equilibrium quantities do
not vanish anywhere and disappear only at infinity one can
introduce the concept of cross section radius $a$ of a force-free
magnetic flux rope. It can be defined as the distance at which the
current flowing in the cylinder can be neglected with sufficient
accuracy. The axial $I_{z}(r)$ and azimuthal $I_{\varphi}(r)$
currents flowing inside a cylinder of radius $r$ and height $2\pi
R$ are:
\begin{equation}
\label{Izz}
I_{z}(r)=2\pi\int_{0}^{r}j_{z}r\,dr=I_{z}\frac{(r/r_{\ast})^{2}}{(1+r^{2}/r_{\ast}^{2})},
\end{equation}
\begin{equation}
\label{Ifi}
I_{\varphi}(r)=2\pi\int_{0}^{r}j_{\varphi}R\,dr=I_{\varphi}\frac{(r/r_{\ast})^{2}}{(1+r^{2}/r_{\ast}^{2})},
\end{equation}
where $I_{\varphi}=I_{z}R/r_{\ast}$ is the total azimuthal
current. It can be seen that although the currents are distributed
to infinity, the main part of them is concentrated near the
$z$-axis. For example, assuming that the magnetic rope is bounded
by a cylindrical surface of radius $a$, from eqs. (\ref{Izz}) and
(\ref{Ifi}) it can be seen that if
\begin{equation}
a=3r_{\ast}=3\sqrt{\frac{I_{z}}{\pi j_{0}}}=\frac{3cB_{0}}{2\pi
j_{0}},
\end{equation}
then about 90\% of the total current flows inside this cylinder.
At distances greater than $a$, the total current and the axial
magnetic field with such accuracy are equal to zero, and the
azimuthal magnetic field decreases as $\sim 1/r$.

From eqs. (\ref{B-phi2}) and (\ref{B-z}) we find the energy stored
inside the force-free magnetic flux rope:
\begin{equation}
E=2\pi\int_{0}^{a}\left(\frac{B_{\varphi}^{2}}{8\pi}+\frac{B_{z}^{2}}{8\pi}\right)r\,dr
=\frac{I_{z}^{2}}{2c^{2}}\ln
\left(1+\frac{a^{2}}{r_{\ast}^{2}}\right).
\end{equation}
It can be seen that the energy is determined by the total axial
current and weakly depends on $a/r_{\ast}$.

In conclusion, we have obtained the integral equilibrium equation
for a cylindrically symmetric force-free magnetic flux rope. Using
this equation, a new local relationship between the axial magnetic
field strength and electric current density was found. This
differs significantly from the Gold-Hoyle work \cite{Gold1960}, as
well as from subsequent works
\cite{Marubashi2007,Vandas2017,Vandas2019,Wang2018} on equilibrium
of a force-free magnetic flux rope, which used some
phenomenological assumptions. As a result, we have derived a new
nonlinear equation (which apparently has never been encountered
before in other physical problems either) for the axial current
density from the equations of magnetostatics without any
phenomenological assumptions, except for the physically natural
assumption that the axial magnetic field decreases at infinity
faster than $1/r$. We have found an analytical solution to this
equation, which made it possible to completely solve the problem
of finding a self-consistent equilibrium of the force-free
magnetic flux rope. We have obtained exact analytical expressions
for the self-consistent components of the electric current density
and magnetic field. It is shown that all equilibrium quantities
are determined by the total axial current (or the amplitude of the
axial magnetic field) and the maximum density of this current. It
was established that the found magnetic field qualitatively
coincides with the Gold-Hoyle magnetic field. However, unlike the
Gold-Hoyle solution, in the solution we obtained, the twist of the
magnetic field lines is not an arbitrary value, but is determined
by the axial current.

The results obtained may be of interest for studies in the solar
atmosphere, solar wind, as well as in other astrophysical and
geophysical problems with low-pressure plasma.

\end{document}